\documentclass{article}
\usepackage[utf8]{inputenc}
\usepackage{amsmath}
\usepackage[paper=a4paper,margin=20mm]{geometry}
\usepackage{graphicx}
\usepackage{authblk}

\title{Dynamics of cholesteric liquid crystal with variable pitch}
\author[1]{I M Tambovtsev}
\author[1,2]{I S Lobanov}
\affil[1]{Department of Physics, St. Petersburg State University, St. Petersburg 198504, Russia}
\affil[2]{Faculty of Physics, ITMO University, 197101 Saint Petersburg, Russia}

\date{April 2022}

\DeclareMathOperator{\Div}{div}
\DeclareMathOperator{\Grad}{grad}
\DeclareMathOperator{\Rot}{rot}
\DeclareMathOperator{\const}{const}
\newcommand{\bn}{\mathbf{n}}
\newcommand{\bx}{\mathbf{x}}
\newcommand{\bH}{\mathbf{H}}
\newcommand{\ba}{\mathbf{a}}

\renewcommand{\epsilon}{\varepsilon}

\begin{document}
\maketitle
\begin{abstract}
    We have written expressions for the free energy of a cholesteric liquid crystal in an approximation using the elasticity constants $K_1, K_2, K_3$ and the energy variation and the corresponding energy and energy gradient along the direction of the director vector $\bn$ at each lattice point for the lattice model. A feature of this lattice model is the shift of the $\bn$ components in space relative to each other. An assessment has been made of the error associated with the respective substitution. Such an approach can increase the accuracy of calculations of the energy and dynamics of localized states in liquid crystals on low-density lattices.
\end{abstract}

\section{Introduction}

The Frank free energy density $\omega$ of liquid crystal (LC) is given by \cite{degennes:book:1993}:
\[
	\omega[\bn]=\frac12\bigg[K_1(\Div\bn)^2 + K_2(\bn\cdot\Rot\bn+q_0)^2 + K_3(\bn\times\Rot\bn)^2-\Delta\chi(\bn\cdot \bH)^2\bigg],
\]
where $K_1$, $K_2$, $K_3$ are Frank constants, 
$q_0=2\pi/P$ is the wavenumber, $P$ is the pitch of the cholesteric helix,
$\bH$ is the external magnetic field
and $\Delta\chi=\chi_{\parallel}-\chi_{\perp}$ is the difference of the magnetic susceptibility $\Delta_{\parallel}$ in the direction of $\bn$
and the one $\Delta_{\perp}$ in the perpendicular direction \cite{knyazev:lc:2013}.
Typical values of the Frank constants are \cite{hakemi:jcp:1983,bogi:lc:2001,bradshaw:mclc:1981,hara:mclc:1985,srivastava:jpcm:2004}
\[
0.5<\frac{K_2}{K_1}<0.8,\quad 0.5<\frac{K_3}{K_1}<3.0.
\]
We assume $K_1,K_2,K_3$ to be fixed.
The wavenumber $q_0$ can be of positive, negative or zero and it is a function of $\bx$.
The value $\Delta\chi$ depends on the properties of the LC and can be positive and negative.

The expression of the energy density can be significantly simplified, if all Frank constants are equal.
It is convenient to have the one constant approximation energy density as explicit addendum
interpreting other contributions as perturbations.

The goal can be achieved applying a sequence of transforms.
The first one 
\[
(\Div\bn)^2+(\Rot\bn)^2+\nabla\cdot((\nabla\cdot \bn)\bn-\bn(\nabla\cdot\bn)) = (\Grad\bn)^2\forall x.
\]
The last addendum on the left hand side describes surface interaction, and can be eliminated by application of the divergence theorem 
and rigid boundary conditions.
The remaining part is read as follows using explicit indices and denoting $n_{i,j}=\partial n_i/\partial x_j$:
\[
\bigg(\sum_{i} n_{i,i}\bigg)^2+\sum_i\bigg(\sum_{j,k} \epsilon_{ijk}n_{k,j}\bigg)^2=\sum_{i,j}(n_{i,j})^2.
\]
The second identity is Pythagorean theorem for the rotor:
\[
\bn^2(\nabla\times\bn)^2=(\bn\cdot(\nabla\times\bn))^2+(\bn\times(\nabla\times\bn))^2\forall x,
\]
where the addenda on the right hand side are squares of projections of rotor on $\bn(x)$ and $\bn(x)^\perp$.
Taking into account the two identities and constrains $\bn(x)^2=1$, 
we write energy density in the following form:
\[
\omega=\frac12\bigg[K_1(\nabla\bn)^2-K_1(\nabla\times\bn)^2+K_2(\bn\cdot\nabla\times\bn)^2+2K_2q_0(\bn\cdot\nabla\times\bn)
+K_2q_0^2
+K_3(\nabla\times\bn)^2-K_3(\bn\cdot(\nabla\times\bn))^2
-\Delta\chi(\bn\cdot \bH)^2
\bigg].
\]
The constant term is of little interest and can be neglected doing
relaxation of the director,
then collecting similar terms the energy density is expressed as follows:
\[
\omega=\frac12\bigg[
K_1(\nabla\bn)^2+(K_3-K_1)(\nabla\times\bn)^2+(K_2-K_3)(\bn\cdot\nabla\times\bn)^2
+2K_2q_0(\bn\cdot\nabla\times\bn)-\Delta\chi(\bn\cdot \bH)^2
\bigg].
\]
The last addendum appears if LC subjected to the external magnetic field \cite{deuling:mclc:1972deformation, cheng:jap:1981}.
The penultimate term describes chiral liquid.
The first term represents energy in one constant approximation,
and the next two terms are non zero only if all three Frank constants are different.
It is worth noting that the expression is quadratic with respect to $\bn$ 
except of the third term $(\bn\cdot\nabla\times\bn)^2$.
If we are only interested in shrinking of cholesteric fingers \cite{oswald:pr:2000,rybakov:prl:2015, tai:pre:2020:smalyukh}, which is controlled by $K_3$,
then we can simplify our life setting $K_2=K_3$ and avoid analysis of non-quadratic energy functional.

\section{Energy variation}

The Frank free energy is just an integral of the density $\omega$ over the volume $V$ of the considered LC: 
\[
E[\bn]=\int_V\omega[\bn]\,d\bx.
\]
The variation $\delta E/\delta\bn$ of the free energy gives direction of relaxation of LC and plays crucial part 
in dynamics of LC:
\[
E[\bn+\epsilon\delta\bn]=E[\bn]+\epsilon\int_V \delta E[\bn]\cdot\delta\bn\,d\bx+o(\epsilon),\text{ as }\epsilon\to0.  
\]
Below we assume implicit dependence of all values on $\bx$,
the notation $a_i$ is used for $i$-th coordinate of a vector $\ba=(a_1,a_2,a_3)$,
and the partial derivatives with respect to components of $\bx=(x_1,x_2,x_3)$
are denoted by comma in the sub-indices:
\[
a_{i,j}=\frac{\partial a_i}{\partial x_j}.
\]
The energy $E$ consist of the several contributions of the form:
\[
F[\bn]=\int_V f(\bn,\Grad\bn)\,d\bx,
\]
for the corresponding contributions $f$ to the energy density $\omega$.
First we derive general rule for variation of a functional $F$ of the given form:
\[
F[\bn+\epsilon\delta\bn]
=\int_V f(\bn+\epsilon\delta\bn,\Grad\bn+\epsilon\Grad\delta\bn)\,d\bx
=F[\bn]+\epsilon\sum_i\int_V \bigg[
\frac{\partial f}{\partial n_i}\delta n_i+\sum_j \frac{\partial f}{\partial n_{i,j}}\delta n_{i,j}
\bigg]d\bx+o(\epsilon).
\]
Further on we drop the sums from the notation, 
assuming summation over repeating indices (regardless of the index position),
unless the position of the sum is unclear.
Integrating by parts and discarding integrals over boundary (the appropriate \emph{boundary conditions}
should be \emph{assumed}), we can allocate the common multiplier $\delta\bn$ under the integral:
\[
F[\bn+\epsilon\delta\bn]-F[\bn]=\epsilon\int_V\bigg[
\frac{\partial f}{\partial n_i}-\frac{\partial}{\partial x_j}\frac{\partial f}{\partial n_{i,j}}
\bigg]\delta n_i\,d\bx+o(\epsilon).
\]
We derived the following formula for the variation of $F$:
\[
\delta F_i=\delta\int_V f dx
=\frac{\partial f}{\partial n_i}
-\sum_j \frac{\partial}{\partial x_j}\frac{\partial f}{\partial n_{i,j}}.
\]
Now we proceed as follows: we separate the addenda in $E$ and compute variation for each of them.
Next we restore constant multipliers and write down the result summing all the contributions.

Strictly speaking we are interested in variation of $\bn$ under constrains $\bn(x)^2=1$,
which translates to constrains on variations of $\delta\bn$:
\[
1=(\bn+\epsilon\delta\bn)^2=\bn^2+2\epsilon\bn\cdot \delta\bn+\epsilon^2\delta\bn^2=1+2\epsilon\bn\cdot \delta\bn+O(\epsilon^2).
\]
Since variations are small, $\epsilon\to 0$, the perturbations of $\bn$ should belong to tangent space to the constrain manifold:
\[
\bn(x)\cdot\delta\bn(x)=0\forall x.
\]
That means that total parallel to $\bn$ component of variation $\delta E$ is never used and can be neglected.
Moreover, dynamics of LC is defined by relaxation, 
that is $d\bn/dt=\dot\bn$ should be anti-parallel to $\delta E[\bn]$ in the tangent to the configuration space.
Since the normalization $\bn^2(x)=1$ should be always preserved,
doing dynamics we have to project $\delta E$ to the tangent space:
\[
\const \dot\bn = \delta E - (\delta E\cdot\bn)\bn\;\forall x.
\]

The energy of the system is defined everywhere, but it has physical meaning only on the constrain manifold $\bn(x)^2=1$.
If the problem is well posed, the dynamics should be independent of definition of energy outside of the manifold.
To check the assumption we consider addendum $\tilde E=(1-\bn^2)G(\bn,\nabla\bn)$ to the energy,
where the square over $\bn$ is taken pointwise $\bn^2(x)=\bn(x)^2$ and the function $G$ is smooth.
The addition of the perturbation $\tilde E$ to the energy describes all sufficiently smooth transformation of the energy,
preserving energy on the constrain manifold.
The variation of the perturbation
\[
\delta \tilde E = -2\bn G+(1-\bn^2)\delta G,
\]
consists of two contributions, the first one vanishes doing projection to the tangent space,
and the second one is zero on the constrain manifold,
therefore the dynamics is unaffected by alteration of the energy outside of the configuration space as expected.
Therefore, the expression of the energy can be simplified using arbitrary formulas
preserving the values for $\bn(x)^2=1$.

Several contributions are squares of simpler quantities.
Let us derive variation for the squares $f(\bn,\nabla\bn)=g(\bn,\nabla\bn)^2$,
where $g$ can be vector valued:
\[
f=g_pg_p,\quad 
\frac{\partial f}{\partial n_i}=2g_p\frac{\partial g_p}{\partial n_i},\quad
\frac{\partial f}{\partial n_{i,j}}=2g_p\frac{\partial g_p}{\partial n_{i,j}},\quad
\frac{\partial}{\partial x_j}\frac{\partial f}{\partial n_{i,j}}
=2\frac{d g_p}{d x_j}\frac{\partial g_p}{\partial n_{i,j}}
+2g_p\frac{\partial }{\partial x_j}\frac{\partial g_p}{\partial n_{i,j}}.
\]
\[
\frac{d g_p}{d x_j}=\sum_i \frac{\partial g_p}{\partial n_i}n_{i,j}
+\sum_{i,k}\frac{\partial g_p}{\partial n_{i,k}}n_{i,kj}.
\]
Hence the variation of the square can be expressed in terms of variation of the function it self:
\[
\delta_i \int_V g^2\,dx
= 2g_p\,\delta_i \int_V g_p\,dx - 2\sum_j\frac{d g_p}{d x_j}\frac{\partial g_p}{\partial n_{i,j}}.
\]

(I) Let $f=\|\nabla\bn\|^2=\sum_{ij} n_{i,j}^2$.
Then 
\[
\frac{\partial f}{\partial n_i}=0,\quad 
\frac{\partial f}{\partial n_{k,j}}=2n_{k,j},\quad
\sum_j \frac{\partial}{\partial x_j}\frac{\partial f}{\partial n_{k,j}}
=2\sum_j n_{k,jj}.
\]
Hence
\[
\delta_k \int_V \|\nabla\bn\|^2\,dx 
= -2\sum_j n_{k,jj},
\]
or in vector calculus notations
\[
\delta \int_V \|\nabla\bn\|^2\,dx 
= -2\Delta \bn.
\]

(II) Let $g=\nabla\times\bn$.
The coordinates of vector product and rotor can be written using Levi-Civita symbol $\epsilon_{ijk}$:
\[
(a\times b)_k = \epsilon_{ijk}a_ib_j = \epsilon_{kij}a_ib_j,\quad
(\nabla\times a)_k = \epsilon_{klm}\frac{\partial}{\partial x_l}a_m = \epsilon_{klm} a_{m,l}.
\]
Also recall the following rule for convolution of the symbols:
\[
\epsilon_{ijk}\epsilon_{imn}=\delta_{jm}\delta_{kn}-\delta_{jn}\delta_{km}.
\]
The first derivatives of the rotor are constant, hence its variation is zero:
\[
\frac{\partial g_p}{\partial n_k} = 0,\quad 
\frac{\partial g_p}{\partial n_{i,j}} = \epsilon_{pji},\quad
\frac{d}{dx_j}\frac{\partial g_p}{\partial n_{i,j}} = 0,\quad 
\Rightarrow\quad
\delta\int_V (\nabla\times\bn)dx = 0.
\]
The variation of its square does not vanish:
\[
\frac{dg_p}{d x_j} 
= \frac{\partial g_p}{\partial n_{i,k}}n_{i,kj} 
= \epsilon_{pkl}n_{l,kj},
\]
\[
\delta_i \int_V (\nabla\times\bn)^2\,dx
= - 2\frac{d g_p}{d x_j}\frac{\partial g_p}{\partial n_{i,j}}
= -2\epsilon_{pkl}n_{l,kj}\epsilon_{pji}
= -2(\delta_{kj}\delta_{li}-\delta_{ki}\delta_{lj})n_{l,kj}
= -2\big(
n_{i,kk}-n_{j,ij}
\big).
\]
The answer in vector calculus  notations takes form:
\[
\delta\int_V (\nabla\times\bn)^2dx 
= -2(\Delta\bn-\Grad\Div\bn).
\]

(III) 
Let $g=\bn\cdot\nabla\times \bn$ or in coordinate form:
\[
g=n_p\epsilon_{plm}n_{m,l}.
\]
This time we deal with quadratic form $g$ of $\bn$, hence the partial derivatives do not vanish:
\[
\frac{\partial g}{\partial n_i} = \epsilon_{ilm}n_{m,l},\quad
\frac{\partial g}{\partial n_{i,j}} = n_p\epsilon_{pji},\quad
\frac{d}{dx_j}\frac{\partial g}{\partial n_{i,j}} = n_{p,j}\epsilon_{pji}.
\]
Hence the variation of $g$ itself is the following linear transform of $\bn$: 
\[
\delta_i\int_V(\bn\cdot\nabla\times\bn)dx 
= \frac{\partial g}{\partial n_i} - \frac{\partial}{\partial x_j}\frac{\partial g}{\partial n_{i,j}}
= \epsilon_{ilm}n_{m,l} - n_{p,j}\epsilon_{pji}
= \epsilon_{ilm}n_{m,l} - n_{m,l}\epsilon_{mli}
= 2\epsilon_{ilm}n_{m,l},
\]
or in vector calculus notation:
\[
\delta\int_V(\bn\cdot\nabla\times\bn)dx 
= 2 \Rot \bn.
\]
Now we are in a position to derive variation of the square of $g$.
Recall that
\[
\delta_i \int_V (\bn\cdot\nabla\times\bn)^2\,dx
= 2g\,\delta_i \int_V g\,dx - 2\frac{d g}{d x_j}\frac{\partial g}{\partial n_{i,j}}.
\]
Using derived above identity, the second addendum can be transformed as follows:
\[
\frac{d g}{d x_j}\frac{\partial g}{\partial n_{i,j}}
= g_{,j}n_p\epsilon_{pji} = (\bn\times\nabla g)_i.
\]
Hence 
\[
\delta \int_V (\bn\cdot\nabla\times\bn)^2\,dx
= 4(\bn\cdot\nabla\times\bn)(\nabla\times\bn)
-2\bn\times\nabla(\bn\cdot\nabla\times\bn).
\]


(IV) Let $f=q g$, $g=\bn\cdot\nabla\times\bn$ and $q$ is a function of $x$ only.
Then
\[
\frac{\partial f}{\partial n_i} = q\frac{\partial g}{\partial n_i},\quad
\frac{\partial f}{\partial n_{i,j}} = q\frac{\partial g}{\partial n_{i,j}},\quad
\frac{d}{d x_j}\frac{\partial f}{\partial n_{i,j}} = q_{,j}\frac{\partial g}{\partial n_{i,j}}+q\frac{d}{d x_j}\frac{\partial g}{\partial n_{i,j}}.
\]
Hence
\[
\delta_i\int_V f dx
=\frac{\partial f}{\partial n_i}-\sum_j \frac{\partial}{\partial x_j}\frac{\partial f}{\partial n_{i,j}}
=q\delta_i\frac12\int_V g dx-q_{,j}\frac{\partial g}{\partial n_{i,j}}.
\]
Using 
\[
\frac{\partial g}{\partial n_{i,j}} = n_p\epsilon_{pji}\quad\Rightarrow\quad
q_{,j}\frac{\partial g}{\partial n_{i,j}}=q_{j,}n_p\epsilon_{pji}=(\bn\times\nabla q)_i,
\]
we conclude that
\[
\delta\int_V (\bn\cdot\nabla\times\bn)q\,dx
=q\delta\int_V (\bn\cdot\nabla\times\bn)\,dx
-\bn\times\nabla q
=2q\Rot \bn
-\bn\times\nabla q.
\]

Finally, variation of energy is as follows:
\[
\delta E 
= -K_1\Delta \bn
-(K_3-K_1)[\Delta\bn-\nabla(\nabla\cdot\bn)]
+(K_2-K_3)[2(\bn\cdot\nabla\times\bn)(\nabla\times\bn)-\bn\times\nabla(\bn\cdot\nabla\times\bn)]
+K_2(2q\Rot \bn-\bn\times\nabla q)
\]
\[
= -K_1\nabla(\nabla\cdot\bn)
-K_3[\Delta\bn-\nabla(\nabla\cdot\bn)]
+(K_2-K_3)[2(\bn\cdot\nabla\times\bn)(\nabla\times\bn)-\bn\times\nabla(\bn\cdot\nabla\times\bn)]
+K_2(2q\nabla\times\bn-\bn\times\nabla q)
\]

\section{Staggered grid discretization}

We assume the discrete square lattice for the numerical calculations with the energy given by the summation over all nodes:

\begin{multline}\label{eq:energy_lattice}
	E=\frac12\sum_{i,j,k}C^{\{ijk\}}\bigg[
K_1(\nabla\bn^{\{ijk\}})^2+(K_3-K_1)(\nabla\times\bn^{\{ijk\}})^2+(K_2-K_3)(\bn^{\{ijk\}}\cdot\nabla\times\bn^{\{ijk\}})^2
+\\+
2K_2q_0(\bn^{\{ijk\}}\cdot\nabla\times\bn^{\{ijk\}})-\Delta\chi(\bn^{\{ijk\}}\cdot \bH^{\{ijk\}})^2,
\end{multline}

\begin{figure}[h]
    \centering
\includegraphics[width=0.5\textwidth]{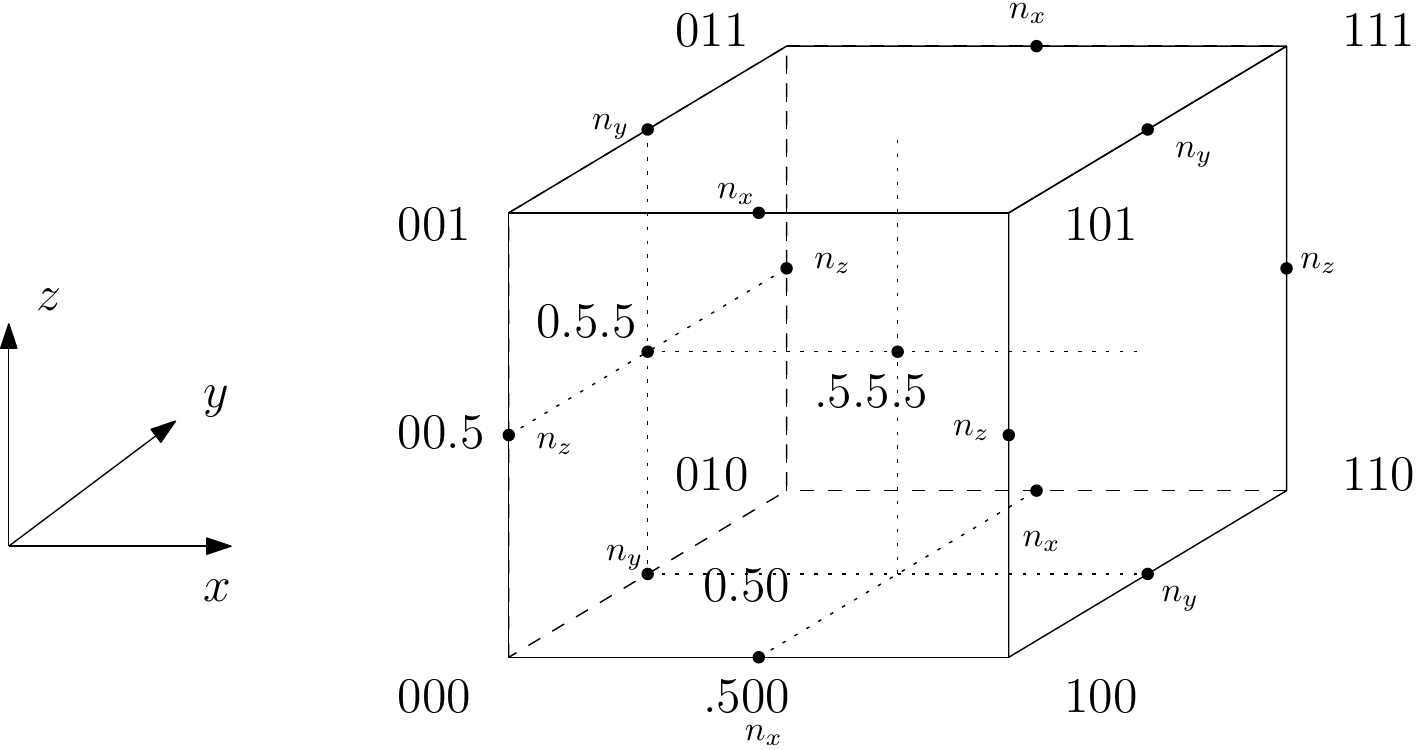}
    \caption{Cross indexing}
    \label{fig:my_label}
\end{figure}

$\{ijk\}$ indices corresponds to the cell position at $x,y,z$ axis respectively. To introduce the differential operations one might need to use the internal indexing inside the cube $\{ijk\}$ (Fig.~\ref{fig:my_label}).

For the convenience of the numerical calculations we introduce the following indexing:
\[
n_x^{i+.5jk}=n_x^{\{ijk\}}, \quad n_y^{ij+.5k}=n_y^{\{ijk\}},\quad n_z^{ijk+.5}=n_z^{\{ijk\}},
\]
\[ n_x^{0.5.5} = (n_x^{.500}+n_x^{.510}+n_x^{.501}+n_x^{.511}+n_x^{-.500}+n_x^{-.510}+n_x^{-.501}+n_x^{-.511})/8, \]
as well as the differential notation:
\[
\partial_y^{0.5.5}n_z=n^{01.5}_z-n^{00.5}_z,
\]
the remaining components are obtained by similar transformations.

The volume of the system is presented as
\[
V= [0, l_xN_x]\times [0, l_yN_y] \times [0, l_zN_z].
\]

Here is a detailed analysis of the error associated with the spacing of lattices $\bn$ at different points for each of its components.

Consider replacing the integral along the x-axis with the corresponding finite sum with coefficients, corresponding to the trapezoid method:
\[
    \int_0^{l_xN_x} f(x)dx 
    = \sum_{i=0}^{N_x} C^i_t f(il_x)l_x+O(l_x^2),\quad l_x\to0, l_xN_x=L_x=\mathrm{const},\quad C^0_t=C^N_t=\frac12,\quad C^j_t=1,j\neq0,N.  
\]
Similarly for the rectangle method:
\[
    \int_0^{l_xN_x} f(x)dx 
    = \sum_{i=0}^{N_x-1} C^i_r f((i+\frac12)l_x)l_x+O(l_x), \quad C^i_r=1.
\]
We repeat the operation for integration over three-dimensional space:
\begin{multline*}
    \iiint_V f(x,y,z)dx\,dy\,dz 
    = 
    \int_0^{l_xN_x}dx \int_0^{l_yN_y}dy \int_0^{l_zN_z}dz f(x,y,z)
    = 
    \int_0^{l_xN_x}dx \int_0^{l_yN_y}dy 
        \left[
            \sum_{k=0}^{N_z} f(x,y,kl_z)C^k_tl_z +O(l_z^2)
        \right]\\
    = 
    \int_0^{l_xN_x}dx
        \left[
            \sum_{j=0}^{N_y} \sum_{k=0}^{N_z}
                f(x,jl_y,kl_z)
                    C^j_tC^k_tl_yl_z
            +O(l_y^2) +O(l_z^2)
        \right] \\
    =
    \sum_{i=0}^{N_x} \sum_{j=0}^{N_y} \sum_{k=0}^{N_z}
        f(il_x,jl_y,kl_z)
            C^i_tC^j_tC^k_tl_xl_yl_z
    + O(l_x^2)+O(l_y^2) +O(l_z^2).
\end{multline*}
Thus, the error of this replacement does not exceed the square of the smallness of the partition along each of the axes. Let us make sure that the differentiation operations do not worsen the accuracy of the approximation:
\[
    (\partial_x n_x)^{ijk}=\frac{n_x^{i+0.5,j,k}-n_x^{i-0.5,j,k}}{l_x}+O(l_x^2),
\]
\begin{multline*}
    \iiint_V (\partial_x n_x)^2 dx\,dy\,dz
    =
    \sum_{i=0}^{N_x} \sum_{j=0}^{N_y} \sum_{k=0}^{N_z}
        \left(
            (\partial_x n_x)^{ijk}
        \right)^2
            C^i_tC^j_tC^k_t
                l_xl_yl_z +O(l_x^2) +O(l_y^2) +O(l_z^2) \\
    =
    \sum_{i=0}^{N_x} \sum_{j=0}^{N_y} \sum_{k=0}^{N_z}
        \left(
            \frac{n_x^{i+0.5,j,k}-n_x^{i-0.5,j,k}}{l_x}+O(l_x^2)
        \right)^2
            C^i_tC^j_tC^k_t
                l_xl_yl_z +O(l_x^2) +O(l_y^2) +O(l_z^2) \\
    =
    \sum_{i=0}^{N_x} \sum_{j=0}^{N_y} \sum_{k=0}^{N_z}
        l_x^{-2}(n_x^{i+0.5,j,k}-n_x^{i-0.5,j,k})^2
            C^i_tC^j_tC^k_tl_xl_yl_z
    +\left(
        \sum_{i=0}^{N_x}O(l_x^2)=O(l_x)
    \right)  
    +O(l_x^2) + O(l_y^2) + O(l_z^2).
\end{multline*}
Similarly, one can show that this order of error is preserved for all such replacements. Such a procedure wad made for all of the energy components. 

(V.1) $(\nabla \bn)^2$ component can be written as the sum:
\begin{multline*}
    \int_V (\nabla \bn)^2 = 
    \int_V
        \left[
            (\partial_x n_x)^2+(\partial_y n_y)^2+(\partial_z n_z)^2
        \right]
    +
    \int_V
        \left[
            (\partial_x n_y)^2+ (\partial_y n_x)^2
        \right]
    +\\+
    \int_V
        \left[
            (\partial_x n_z)^2+ (\partial_z n_x)^2
        \right]
    +\int_V
        \left[
            (\partial_y n_z)^2+ (\partial_z n_y)^2
        \right]
    ,
\end{multline*}
and all of them can be rewritten in a new form:
\[
    \int_V
    \left[
        (\partial_x n_x)^2+(\partial_y n_y)^2+(\partial_z n_z)^2
    \right]
    =
    \sum_{i=0}^{N_x} \sum_{j=0}^{N_y} \sum_{k=0}^{N_z}
        C_t^iC_t^jC_t^k
            \left[
                (\partial_x n_x)^2+(\partial_y n_y)^2+(\partial_z n_z)^2
            \right]^{ijk}
    ,
\]
\[
    \int_V 
    \left[
        (\partial_x n_y)^2+ (\partial_y n_x)^2
    \right] 
    =
    \sum_{i=0}^{N_x-1} \sum_{j=0}^{N_y-1} \sum_{k=0}^{N_z}
        C_r^i C_r^j C_t^k
            \left[
                (\partial_x n_y)^2+ (\partial_y n_x)^2
            \right]^{i+.5j+.5k}
    ,
\]
\[
    \int_V
        \left[
            (\partial_x n_z)^2+ (\partial_z n_x)^2
        \right]
    =
    \sum_{i=0}^{N_x-1} \sum_{j=0}^{N_y} \sum_{k=0}^{N_z-1}
        C_r^i C_t^j C_r^k
            \left[
                (\partial_x n_z)^2+ (\partial_z n_x)^2
            \right]^{i+.5jk+.5}
    ,
\]
\[
    \int_V
        \left[
            (\partial_y n_z)^2+ (\partial_z n_y)^2
        \right]
    =
    \sum_{i=0}^{N_x} \sum_{j=0}^{N_y-1} \sum_{k=0}^{N_z-1}
        C_t^i C_r^j C_r^k
            \left[
                (\partial_y n_z)^2+ (\partial_z n_y)^2
            \right]^{ij+.5k+.5}
    .
\]
The resulting expressions can be written in terms of the values of $\bn$ on the lattice:
\[
    \left[
        (\partial_x n_x)^2+(\partial_y n_y)^2+(\partial_z n_z)^2
    \right]^{ijk}
    =
    \left[ 
        \left(n_x^{\{ijk\}}-n_x^{\{i-1jk\}}\right)^2+
        \left(n_y^{\{ijk\}}-n_y^{\{ij-1k\}}\right)^2+
        \left(n_z^{\{ijk\}}-n_z^{\{ijk-1\}}\right)^2
    \right]
    ,
\]
\[
    \left[
        (\partial_x n_y)^2+ (\partial_y n_x)^2
    \right]^{i+.5j+.5k}
    =
    (n_y^{\{i+1jk\}}-n_y^{\{ijk\}})^2 + (n_x^{\{ij+1k\}}-n_x^{\{ijk\}})^2
    ,
\]
\[
    \left[
        (\partial_x n_z)^2+ (\partial_z n_x)^2
    \right]^{i+.5jk+.5}
    =
    (n_z^{\{i+1jk\}}-n_z^{\{ijk\}})^2 + (n_x^{\{ijk+1\}}-n_x^{\{ijk\}})^2
    ,
\]
\[
    \left[
        (\partial_y n_z)^2+ (\partial_z n_y)^2
    \right]^{ij+.5k+.5}
    =
    (n_z^{\{ij+1k\}}-n_z^{\{ijk\}})^2 + (n_y^{\{ijk+1\}}n_y^{\{ijk\}})^2
    .
\]
(V.2) Operation $(\nabla \times \bn)^2$ can be written in each grid cell as: 
\[
    \nabla \times \bn
    =
    \begin{bmatrix}
        \partial_y^{0.5.5}n_z-\partial_z^{0.5.5}n_y\\
        \partial_z^{.50.5}n_x-\partial_x^{.50.5}n_z\\
        \partial_x^{.5.50}n_y-\partial_y^{.5.50}n_x
    \end{bmatrix}
    .
\]
Then, the $(\nabla \times \bn)^2$ term of the energy may be written as sum:
\begin{multline*}
    \int_V(\nabla \times \bn)^2 = 
    \sum_{i=0}^{N_x} \sum_{j=0}^{N_y-1} \sum_{k=0}^{N_z-1}
        C_t^i C_r^j C_r^k
            \left[
                (\partial_y n_z-\partial_z n_y)^2
            \right]^{ij+.5k+.5}
    +\\+
    \sum_{i=0}^{N_x-1} \sum_{j=0}^{N_y} \sum_{k=0}^{N_z-1}
            C_r^i C_t^j C_r^k
        \left[
            (\partial_z n_x-\partial_x n_z)^2
        \right]^{i+.5jk+.5}
    +
    \sum_{i=0}^{N_x-1} \sum_{j=0}^{N_y-1} \sum_{k=0}^{N_z}
        C_r^i C_r^j C_t^k
            \left[
                (\partial_x n_y-\partial_y n_x)^2
            \right]^{i+.5j+.5k}
    ,
\end{multline*}
and the resulting expressions can be written in terms of the values of $\bn$ on the lattice:
\[
    \left[
        (\partial_y n_z-\partial_z n_y)^2
    \right]^{ij+.5k+.5}=
    \left[
        (n_z^{\{ij+1k\}}-n_z^{\{ijk\}})-(n_y^{\{ijk+1\}}-n_y^{\{ijk\}})
    \right]^2
    ,
\]
\[
    \left[
        (\partial_z n_x-\partial_x n_z)^2
    \right]^{i+.5jk+.5}
    =
    \left[
        (n_x^{\{ijk+1\}}-n_x^{\{ijk\}})-(n_z^{\{i+1jk\}})-n_z^{\{ijk\}})
    \right]^2,
\]
\[
    \left[
        (\partial_x n_y-\partial_y n_x)^2
    \right]^{i+.5j+.5k}
    =
    \left[
        (n_y^{\{i+1jk\}}-n_y^{\{ijk\}})- (n_x^{\{ij+1k\}}-n_x^{\{ijk\}})
    \right]^2
    .
\]
(V.3) Operation $\bn \cdot (\nabla \times \bn)$ can be written in each grid cell as: 
\[ 
    \bn \cdot (\nabla \times \bn)
    =
    \begin{bmatrix}
        n_x^{0.5.5}
    \\
        n_y^{.50.5}
    \\
        n_z^{.5.50}
    \end{bmatrix}
        \begin{bmatrix}
            \partial_y^{0.5.5}n_z-\partial_y^{0.5.5}n_z
        \\
            \partial_z^{.50.5}n_x-\partial_x^{.50.5}n_z
        \\
            \partial_x^{.5.50}n_y-\partial_y^{.5.50}n_x
        \end{bmatrix} 
\]
Then, the $\bn \cdot (\nabla \times \bn)$ term of the energy may be written as sum:
\begin{multline*}
    \int_V \bn \cdot (\nabla \times \bn) = 
    \sum_{i=0}^{N_x} \sum_{j=0}^{N_y-1} \sum_{k=0}^{N_z-1}
        C_t^i C_r^j C_r^k
            \left[
                n_x(\partial_y n_z-\partial_y n_z)
            \right]^{ij+.5k+.5}
    +\\+
    \sum_{i=0}^{N_x-1} \sum_{j=0}^{N_y} \sum_{k=0}^{N_z-1}
        C_r^i C_t^j C_r^k
            \left[
                n_y (\partial_z n_x-\partial_x n_z)
            \right]^{i+.5jk+.5}
    +
    \sum_{i=0}^{N_x-1} \sum_{j=0}^{N_y-1} \sum_{k=0}^{N_z}
        C_r^i C_r^j C_t^k
            \left[
                n_z (\partial_x n_y-\partial_y n_x)
            \right]^{i+.5j+.5z}
    ,
\end{multline*}
and the resulting expressions can be written in terms of the values of $\bn$ on the lattice:
\begin{multline*}
    \left[
        n_x(\partial_y n_z-\partial_y n_z)
    \right]^{ij+.5k+.5}
    =
    \frac{1}{8}
        \Big[
            n_x^{\{ijk\}}+n_x^{\{ij+1k\}}+n_x^{\{ijk+1\}}+n_x^{\{ij+1k+1\}}
            +\\+
            n_x^{\{i-1jk\}}+n_x^{\{i-1j+1k\}}+n_x^{\{i-1jk+1\}}+n_x^{\{i-1j+1k+1\}}
        \Big]
            \left[
                (n_z^{\{ij+1k\}}-n_z^{\{ijk\}})-(n_y^{\{ijk+1\}}-n_y^{\{ijk\}})
            \right]
    ,
\end{multline*}
\begin{multline*}
    \left[
        n_y (\partial_z n_x-\partial_x n_z)
    \right]^{i+.5jk+.5}
    =
    \frac{1}{8}
        \Big[
            n_y^{\{ijk\}}+n_y^{\{i+1jk\}}+n_y^{\{ijk+1\}}+n_y^{\{i+1jk+1\}}
            +\\+
            n_y^{\{ij-1k\}}+n_y^{\{i+1j-1k\}}+n_y^{\{ij-1k+1\}}+n_y^{\{i+1j-1k+1\}}
        \Big]
            \left[
                (n_x^{\{ijk+1\}}-n_x^{\{ijk\}})-(n_z^{\{i+1jk\}}-n_z^{\{ijk\}})
            \right]
    ,
\end{multline*}
\begin{multline*}
    \left[
        n_z (\partial_x n_y-\partial_y n_x)
    \right]^{i+.5j+.5z}
    =
    \frac{1}{8}
        \Big[
            n_z^{\{ijk\}}+n_z^{\{ij+1k\}}+n_z^{\{i+1jk\}}+n_z^{\{i+1j+1k\}}
            +\\+
            n_z^{\{ijk-1\}}+n_z^{\{ij+1k-1\}}+n_z^{\{i+1jk-1\}}+n_z^{\{i+1j+1k-1\}}
        \Big]
            \left[
                (n_y^{\{i+1jk\}}-n_y^{\{ijk\}})-(n_x^{\{ij+1k\}}-n_x^{\{ijk\}})
            \right]
    .
\end{multline*}

(V.4) The $(\bn\cdot \bH)^2$ term of the energy may be written as sum: 
\begin{multline*}
    \int_V(\bn\cdot \bH)^2 =
    \sum_{i=0}^{N_x-1} \sum_{j=0}^{N_y} \sum_{k=0}^{N_z}
        C_r^i C_t^j C_t^k
            \left[
                (n_x H_x)^2
            \right]^{i+.5jk}
    +\\+
    \sum_{i=0}^{N_x} \sum_{j=0}^{N_y-1} \sum_{k=0}^{N_z}
        C_t^i C_r^j C_t^k
            \left[
                (n_yH_y)^2
            \right]^{ij+.5k}
    +
    \sum_{i=0}^{N_x} \sum_{j=0}^{N_y} \sum_{k=0}^{N_z-1}
        C_t^i C_t^j C_r^k
            \left[
                (n_z H_z)^2
            \right]^{ijk+.5}
    ,
\end{multline*}
and the resulting expressions can be written in terms of the values of $\bn$ on the lattice:
\[
    \left[
        (n_x H_x)^2
    \right]^{i+.5jk}
    =
    \left(
        n_x^{\{ijk\}} H_x^{\{ijk\}}
    \right)^2,
\]
\[
    \left[
        (n_y H_y)^2
    \right]^{ij+.5k}
    =
    \left(
        n_y^{\{ijk\}} H_y^{\{ijk\}}
    \right)^2
    ,
\]
\[
    \left[
        (n_z H_z)^2
    \right]^{ijk+.5}
    =
    \left(
        n_z^{\{ijk\}} H_z^{\{ijk\}}
    \right)^2
    .
\]

\subsection{Gradient of energy}

Consider the gradient in $\bn$ in the lattice model as an analogue of the energy variation in the continuous model. To write such an expression, we take the gradient from each of the energy components in the lattice model in the direction $\bn$.

(VI.1) Consider the gradient of each component $(\nabla \bn)^2$ at a lattice site $\{ijk\}$:
\begin{multline*}
    \partial_{\bn^{\{ijk\}}}
        \left[
            \sum_{j=0}^{N_y} \sum_{k=0}^{N_z}
                C_t^iC_t^jC_t^k
                    \left[
                            (n_x^{\{ijk\}}-n_x^{\{i-1jk\}})^2+
                            (n_y^{\{ijk\}}-n_y^{\{ij-1k\}})^2+
                            (n_z^{\{ijk\}}-n_z^{\{ijk-1\}})^2
                    \right]
        \right]
    =\\=
    2
    \begin{bmatrix}
        C_t^i C_t^{j} C_t^k
            (n_x^{\{ijk\}}-n_x^{\{i-1jk\}})
        -
        C_r^{i+1} C_r^j C_t^k
            (n_x^{\{i+1jk\}}-n_x^{\{ijk\}})
        \\
        C_t^i C_t^{j} C_t^k
            (n_y^{\{ijk\}}-n_y^{\{ij-1k\}})
        -
        C_r^i C_r^{j+1} C_t^k
            (n_y^{\{ij+1k\}}-n_y^{\{ijk\}})
        \\
        C_t^i C_t^{j} C_t^k
            (n_z^{\{ijk\}}-n_z^{\{ijk-1\}})
        -
        C_r^i C_r^j C_t^{k+1}
            (n_z^{\{ijk+1\}}-n_z^{\{ijk\}})
    \end{bmatrix},
\end{multline*}
\begin{multline*}
    \partial_{\bn^{\{ijk\}}}
        \left[
            \sum_{i=0}^{N_x-1} \sum_{j=0}^{N_y-1} \sum_{k=0}^{N_z}
                C_r^i C_r^j C_t^k
                    \left[
                        (n_y^{\{i+1jk\}}-n_y^{\{ijk\}})^2 + (n_x^{\{ij+1k\}}-n_x^{\{ijk\}})^2
                    \right]
        \right]
    =\\=
    2
        \begin{bmatrix}
            C_r^i C_r^{j-1} C_t^k
                (n_x^{\{ijk\}}-n_x^{\{ij-1k\}})
            -
            C_r^i C_r^j C_t^k
                (n_x^{\{ij+1k\}}-n_x^{\{ijk\}})
        \\
            C_r^{i-1} C_r^{j} C_t^k
                (n_y^{\{ijk\}}-n_y^{\{i-1jk\}})
            -
                C_r^i C_r^j C_t^k
                    (n_y^{\{i+1jk\}}-n_y^{\{ijk\}})
        \\
            0
        \end{bmatrix},
\end{multline*}
\begin{multline*}
    \partial_{\bn^{\{ijk\}}}
        \left[
            \sum_{i=0}^{N_x-1} \sum_{j=0}^{N_y} \sum_{k=0}^{N_z-1}
                C_r^i C_t^j C_r^k
                    \left[
                        (n_z^{\{i+1jk\}}-n_z^{\{ijk\}})^2+ (n_x^{\{ijk+1\}}-n_x^{\{ijk\}})^2
                    \right]
        \right]
    =\\=
    2
        \begin{bmatrix}
            C_r^i C_t^{j} C_r^{k-1}
                (n_x^{\{ijk\}}-n_x^{\{ijk-1\}})
            -
            C_r^i C_t^j C_r^k
                (n_x^{\{ijk+1\}}-n_x^{\{ijk\}})
        \\
            0
        \\
            C_r^{i-1} C_t^{j} C_r^k
                (n_z^{\{ijk\}}-n_z^{\{i-1jk\}})
            -
            C_r^i C_t^j C_r^k
                (n_z^{\{i+1jk\}}-n_z^{\{ijk\}})
        \end{bmatrix}
    ,
\end{multline*}
\begin{multline*}
    \partial_{\bn^{\{ijk\}}}\left[\sum_{i=0}^{N_x} \sum_{j=0}^{N_y-1} \sum_{k=0}^{N_z-1}C_t^i C_r^j C_r^k[(n_z^{\{ij+1k\}}-n_z^{\{ijk\}})^2+ (n_y^{\{ijk+1\}}n_y^{\{ijk\}})^2]\right]
    =\\=
    2
    \begin{bmatrix}
        0
    \\
        C_t^i C_r^{j} C_r^{k-1}
            (n_y^{\{ijk\}}-n_y^{\{ijk-1\}})
        -
        C_t^i C_r^j C_r^k
            (n_y^{\{ijk+1\}}-n_y^{\{ijk\}})
    \\
        C_t^{i} C_r^{j-1} C_r^k
            (n_z^{\{ijk\}}-n_z^{\{ij-1k\}})
        -
        C_t^i C_r^j C_r^k
            (n_z^{\{ij+1k\}}-n_z^{\{ijk\}})
    \end{bmatrix}.
\end{multline*}

(VI.2) Consider the gradient of each component $(\nabla \times \bn)^2$ at a lattice site $\{ijk\}$: 
\begin{multline*}
    \partial_{\bn^{\{ijk\}}}
        \left[
            \sum_{i=0}^{N_x} \sum_{j=0}^{N_y-1} \sum_{k=0}^{N_z-1}
                C_t^i C_r^j C_r^k
                    \left[
                        (n_z^{\{ij+1k\}}-n_z^{\{ijk\}})-(n_y^{\{ijk+1\}}-n_y^{\{ijk\}})
                    \right]^2
        \right]
=\\=
    2
    \begin{bmatrix}
        0
    \\
        C_t^i C_r^j C_r^k
            \left[
                (n_z^{\{ij+1k\}}-n_z^{\{ijk\}})-(n_y^{\{ijk+1\}}-n_y^{\{ijk\}})
            \right]
        -
        C_t^i C_r^j C_r^{k-1}
            \left[
                (n_z^{\{ij+1k-1\}}-n_z^{\{ijk-1\}})-(n_y^{\{ijk\}}-n_y^{\{ijk-1\}})
            \right]
    \\
        -
        C_t^i C_r^j C_r^k
            \left[
                (n_z^{\{ij+1k\}}-n_z^{\{ijk\}})-(n_y^{\{ijk+1\}}-n_y^{\{ijk\}})
            \right]
        +
        C_t^i C_r^{j-1} C_r^k
            \left[
                (n_z^{\{ijk\}}-n_z^{\{ij-1k\}})-(n_y^{\{ij-1k+1\}}-n_y^{\{ij-1k\}})
            \right]
    \end{bmatrix},
\end{multline*}
\begin{multline*}
    \partial_{\bn^{\{ijk\}}}
    \left[
        \sum_{i=0}^{N_x-1} \sum_{j=0}^{N_y} \sum_{k=0}^{N_z-1}
        C_r^i C_t^j C_r^k
        \left[
            (n_x^{\{ijk+1\}}-n_x^{\{ijk\}})-(n_z^{\{i+1jk\}}-n_z^{\{ijk\}})
        \right]^2
    \right]
=\\=
    2
    \begin{bmatrix}
        -
        C_r^i C_t^j C_r^k
            \left[
                (n_x^{\{ijk+1\}}-n_x^{\{ijk\}})-(n_z^{\{i+1jk\}}-n_z^{\{ijk\}})
            \right]
        +
        C_r^i C_t^j C_r^{k-1}
            \left[
                (n_x^{\{ijk\}}-n_x^{\{ijk-1\}})-(n_z^{\{i+1jk-1\}}-n_z^{\{ijk-1\}})
            \right]
    \\
        0
    \\
        C_r^i C_t^j C_r^k
                \left[
                    (n_x^{\{ijk+1\}}-n_x^{\{ijk\}})-(n_z^{\{i+1jk\}}-n_z^{\{ijk\}})
                \right]
        +
        C_r^{i-1} C_t^j C_r^k
            \left[
                (n_x^{\{i-1jk+1\}}-n_x^{\{i-1jk\}})-(n_z^{\{ijk\}}-n_z^{\{i-1jk\}})
            \right]
    \end{bmatrix},
\end{multline*}
\begin{multline*}
    \partial_{\bn^{\{ijk\}}}
        \left[
            \sum_{i=0}^{N_x-1} \sum_{j=0}^{N_y-1} \sum_{k=0}^{N_z}
                C_r^i C_r^j C_t^k
                    \left[
                        (n_y^{\{i+1jk\}}-n_y^{\{ijk\}}) - (n_x^{\{ij+1k\}}-n_x^{\{ijk\}})
                    \right]^2
        \right]
=\\=
    2
    \begin{bmatrix}
        C_r^i C_r^j C_t^k
            \left[
                (n_y^{\{i+1jk\}}-n_y^{\{ijk\}}) - (n_x^{\{ij+1k\}}-n_x^{\{ijk\}})
            \right]
        -
        C_r^i C_r^{j-1} C_t^k
            \left[
                (n_y^{\{i+1j-1k\}}-n_y^{\{ij-1k\}})- (n_x^{\{ijk\}}-n_x^{\{ij-1k\}})
            \right]
    \\
        -
        C_r^i C_r^j C_t^k
            \left[
                (n_y^{\{i+1jk\}}-n_y^{\{ijk\}})- (n_x^{\{ij+1k\}}-n_x^{\{ijk\}})
            \right]
            +
            C_r^{i-1} C_r^j C_t^k
            \left[
                (n_y^{\{ijk\}}-n_y^{\{i-1jk\}})- (n_x^{\{i-1j+1k\}}-n_x^{\{i-1jk\}})
            \right]
    \\
        0
    \end{bmatrix}.
\end{multline*}

\begin{multline*}
    \partial_{\bn^{\{ijk\}}}
    \left[
        \sum_{i=0}^{N_x-1} \sum_{j=0}^{N_y} \sum_{k=0}^{N_z-1}
            C_r^i C_t^j C_r^k
                \left[
                    (n_x^{\{ijk+1\}}-n_x^{\{ijk\}})-(n_z^{\{i+1jk\}}-n_z^{\{ijk\}})
                \right]^2
    \right]
    =\\=
    2
    \begin{bmatrix}
        -
        C_r^i C_t^j C_r^k
            \left[
                (n_x^{\{ijk+1\}}-n_x^{\{ijk\}})-(n_z^{\{i+1jk\}}-n_z^{\{ijk\}})
            \right]
        +
        C_r^i C_t^j C_r^{k-1}
            \left[
                (n_x^{\{ijk\}}-n_x^{\{ijk-1\}})-(n_z^{\{i+1jk-1\}}-n_z^{\{ijk-1\}})
            \right]
\\
        0
\\
        C_r^i C_t^j C_r^k
            \left[
                (n_x^{\{ijk+1\}}-n_x^{\{ijk\}})-(n_z^{\{i+1jk\}}-n_z^{\{ijk\}})
            \right]
        +
        C_r^{i-1} C_t^j C_r^k
            \left[
                (n_x^{\{i-1jk+1\}}-n_x^{\{i-1jk\}})-(n_z^{\{ijk\}}-n_z^{\{i-1jk\}})
            \right]
    \end{bmatrix},
\end{multline*}
\begin{multline*}
    \partial_{\bn^{\{ijk\}}}
    \left[
        \sum_{i=0}^{N_x-1} \sum_{j=0}^{N_y-1} \sum_{k=0}^{N_z}
            C_r^i C_r^j C_t^k
            \left[
                (n_y^{\{i+1jk\}}-n_y^{\{ijk\}}) - (n_x^{\{ij+1k\}}-n_x^{\{ijk\}})
            \right]^2
    \right]
    =\\=
    2
    \begin{bmatrix}
        C_r^i C_r^j C_t^k
            \left[
                (n_y^{\{i+1jk\}}-n_y^{\{ijk\}}) - (n_x^{\{ij+1k\}}-n_x^{\{ijk\}})
            \right]
        -
        C_r^i C_r^{j-1} C_t^k
            \left[
                (n_y^{\{i+1j-1k\}}-n_y^{\{ij-1k\}}) - (n_x^{\{ijk\}}-n_x^{\{ij-1k\}})
            \right]
\\
        -
        C_r^i C_r^j C_t^k
            \left[
                (n_y^{\{i+1jk\}}-n_y^{\{ijk\}})- (n_x^{\{ij+1k\}}-n_x^{\{ijk\}})
            \right]
        +
        C_r^{i-1} C_r^j C_t^k
            \left[
                (n_y^{\{ijk\}}-n_y^{\{i-1jk\}})- (n_x^{\{i-1j+1k\}}-n_x^{\{i-1jk\}})
            \right]
\\
        0
    \end{bmatrix}.
\end{multline*}

(VI.3) Consider each of the components of the gradient of each component $\bn \cdot (\nabla \times \bn)$ at a lattice site $\{ijk\}$: 

\begin{multline*}
    \partial_{n^{\{ijk\}}_x}
        \Bigg[
            \frac{1}{8}
                \sum_{i=0}^{N_x} \sum_{j=0}^{N_y-1} \sum_{k=0}^{N_z-1}
                        C_t^i C_r^j C_r^k
                            \Big[
                                n_x^{\{ijk\}}+
                                n_x^{\{ij+1k\}}+
                                n_x^{\{ijk+1\}}+
                                n_x^{\{ij+1k+1\}}
+\\+
                                n_x^{\{i-1jk\}}+
                                n_x^{\{i-1j+1k\}}+
                                n_x^{\{i-1jk+1\}}+
                                n_x^{\{i-1j+1k+1\}}
                            \Big]
                                \left[
                                    (n_z^{\{ij+1k\}}-n_z^{\{ijk\}})-(n_y^{\{ijk+1\}}-n_y^{\{ijk\}})
                                \right]
        \Bigg]
=\\=
            \frac{1}{8}
                \bigg[
                    C_t^i C_r^j C_r^k
                        \left[
                            (n_z^{\{ij+1k\}}-n_z^{\{ijk\}})-(n_y^{\{ijk+1\}}-n_y^{\{ijk\}})
                        \right]
                    +\\+
                    C_t^i C_r^{j-1} C_r^k
                        \left[
                            (n_z^{\{ijk\}}-n_z^{\{ij-1k\}})-(n_y^{\{ij-1k+1\}}-n_y^{\{ij-1k\}})
                        \right]
                    +\\+
                    C_t^i C_r^j C_r^{k-1}
                        \left[
                            (n_z^{\{ij+1k-1\}}-n_z^{\{ijk-1\}})-(n_y^{\{ijk\}}-n_y^{\{ijk-1\}})
                        \right]
                    +\\+
                    C_t^i C_r^{j-1} C_r^{k-1}
                        \left[
                            (n_z^{\{ijk-1\}}-n_z^{\{ij-1k-1\}})-(n_y^{\{ij-1k\}}-n_y^{\{ij-1k-1\}})
                        \right]
                    +\\+
                    C_t^{i+1} C_r^j C_r^k
                        \left[
                            (n_z^{\{i+1j+1k\}}-n_z^{\{i+1jk\}})-(n_y^{\{i+1jk+1\}}-n_y^{\{i+1jk\}})
                        \right]
                    +\\+
                    C_t^{i+1} C_r^{j-1} C_r^k
                        \left[
                            (n_z^{\{i+1jk\}}-n_z^{\{i+1j-1k\}})-(n_y^{\{i+1j-1k+1\}}-n_y^{\{i+1j-1k\}})
                        \right]
                    +\\+
                    C_t^{i+1} C_r^j C_r^{k-1}
                        \left[
                            (n_z^{\{i+1j+1k-1\}}-n_z^{\{i+1jk-1\}})-(n_y^{\{i+1jk\}}-n_y^{\{i+1jk-1\}})
                        \right]
                    +\\+
                    C_t^{i+1} C_r^{j-1} C_r^{k-1}
                        \left[
                            (n_z^{\{i+1jk-1\}}-n_z^{\{i+1j-1k-1\}})-(n_y^{\{i+1j-1k\}}-n_y^{\{i+1j-1k-1\}})
                        \right]
                \bigg ]
    ,
\end{multline*}

\begin{multline*}
    \partial_{n^{\{ijk\}}_y}
    \Bigg[
        \frac{1}{8}
            \sum_{i=0}^{N_x} \sum_{j=0}^{N_y-1} \sum_{k=0}^{N_z-1}
                C_t^i C_r^j C_r^k
                    \Big[
                        n_x^{\{ijk\}}+n_x^{\{ij+1k\}}+n_x^{\{ijk+1\}}+n_x^{\{ij+1k+1\}}
                        +\\+
                        n_x^{\{i-1jk\}}+n_x^{\{i-1j+1k\}}+n_x^{\{i-1jk+1\}}+n_x^{\{i-1j+1k+1\}}
                    \Big]
                    \left[
                        (n_z^{\{ij+1k\}}-n_z^{\{ijk\}})-(n_y^{\{ijk+1\}}-n_y^{\{ijk\}})
                    \right]
    \Bigg]
=\\=
    -\frac{1}{8}
        \bigg[
            C_t^i C_r^j C_r^{k-1}
            \left[
                n_x^{\{ijk-1\}}+
                n_x^{\{ij+1k-1\}}+
                n_x^{\{ijk\}}+
                n_x^{\{ij+1k\}}+
                n_x^{\{i-1jk-1\}}+
                n_x^{\{i-1j+1k-1\}}+
                n_x^{\{i-1jk\}}+
                n_x^{\{i-1j+1k\}}
            \right]
-\\-
            C_t^i C_r^j C_r^k
                \left[
                    n_x^{\{ijk\}}+
                    n_x^{\{ij+1k\}}+
                    n_x^{\{ijk+1\}}+
                    n_x^{\{ij+1k+1\}}+
                    n_x^{\{i-1jk\}}+
                    n_x^{\{i-1j+1k\}}+
                    n_x^{\{i-1jk+1\}}+
                    n_x^{\{i-1j+1k+1\}}
                \right]
        \bigg]
    ,
\end{multline*}

\begin{multline*}
    \partial_{n^{\{ijk\}}_z}
    \Bigg[
        \frac{1}{8}
            \sum_{i=0}^{N_x} \sum_{j=0}^{N_y-1} \sum_{k=0}^{N_z-1}
                C_t^i C_r^j C_r^k
                    \Big[
                        n_x^{\{ijk\}}+n_x^{\{ij+1k\}}+n_x^{\{ijk+1\}}+n_x^{\{ij+1k+1\}}
+\\+
                        n_x^{\{i-1jk\}}+n_x^{\{i-1j+1k\}}+n_x^{\{i-1jk+1\}}+n_x^{\{i-1j+1k+1\}}
                    \Big]
                        \left[
                            (n_z^{\{ij+1k\}}-n_z^{\{ijk\}})-(n_y^{\{ijk+1\}}-n_y^{\{ijk\}})
                        \right]
    \Bigg]
=\\=
    \frac{1}{8}
        \bigg[
            C_t^i C_r^{j-1} C_r^k
                \Big[
                    n_x^{\{ij-1k\}}+
                    n_x^{\{ijk\}}+
                    n_x^{\{ij-1k+1\}}+
                    n_x^{\{ijk+1\}}+
                    n_x^{\{i-1j-1k\}}+
                    n_x^{\{i-1j-1k\}}+
                    n_x^{\{i-1j-1k+1\}}+
                    n_x^{\{i-1jk+1\}}
                \Big]
            -\\-
            C_t^i C_r^j C_r^k
                \Big[
                    n_x^{\{ijk\}}+
                    n_x^{\{ij+1k\}}+
                    n_x^{\{ijk+1\}}+
                    n_x^{\{ij+1k+1\}}+
                    n_x^{\{i-1jk\}}+
                    n_x^{\{i-1j+1k\}}+
                    n_x^{\{i-1jk+1\}}+
                    n_x^{\{i-1j+1k+1\}}
                \Big]
        \bigg]
    ,
\end{multline*}

\begin{multline*}
    \partial_{n^{\{ijk\}}_x}
    \Bigg[
        \frac{1}{8}
            \sum_{i=0}^{N_x-1} \sum_{j=0}^{N_y} \sum_{k=0}^{N_z-1}
                C_r^i C_t^j C_r^k
                    \Big[
                        n_y^{\{ijk\}}+n_y^{\{i+1jk\}}+n_y^{\{ijk+1\}}+n_y^{\{i+1jk+1\}}
+\\+
                        n_y^{\{ij-1k\}}+n_y^{\{i+1j-1k\}}+n_y^{\{ij-1k+1\}}+n_y^{\{i+1j-1k+1\}}
                    \Big]
                    \left[
                        (n_x^{\{ijk+1\}}-n_x^{\{ijk\}})-(n_z^{\{i+1jk\}})-n_z^{\{ijk\}})
                    \right]
    \Bigg]
=\\=
    \frac{1}{8}
        \bigg[
            C_r^i C_t^j C_r^{k-1}
            \left[
                n_y^{\{ijk-1\}}+
                n_y^{\{i+1jk-1\}}+
                n_y^{\{ijk\}}+
                n_y^{\{i+1jk\}}+
                n_y^{\{ij-1k-1\}}+
                n_y^{\{i+1j-1k-1\}}+
                n_y^{\{ij-1k\}}+
                n_y^{\{i+1j-1k\}}
            \right]
            -\\-
            C_r^i C_t^j C_r^k
            \left[
                n_y^{\{ijk\}}+
                n_y^{\{i+1jk\}}+
                n_y^{\{ijk+1\}}+
                n_y^{\{i+1jk+1\}}+
                n_y^{\{ij-1k\}}+
                n_y^{\{i+1j-1k\}}+
                n_y^{\{ij-1k+1\}}+
                n_y^{\{i+1j-1k+1\}}
            \right]
        \bigg]
,
\end{multline*}
\begin{multline*}
    \partial_{n^{\{ijk\}}_y}
    \Bigg[
        \frac{1}{8}
            \sum_{i=0}^{N_x-1} \sum_{j=0}^{N_y} \sum_{k=0}^{N_z-1}
                C_r^{i} C_t^{j} C_r^{k}
                    \Big[
                        n_y^{\{ijk\}}+n_y^{\{i+1jk\}}+n_y^{\{ijk+1\}}+n_y^{\{i+1jk+1\}}
+\\+                        
                        n_y^{\{ij-1k\}}+n_y^{\{i+1j-1k\}}+n_y^{\{ij-1k+1\}}+n_y^{\{i+1j-1k+1\}}
                    \Big]
                    \left[
                        (n_x^{\{ijk+1\}}-n_x^{\{ijk\}})-(n_z^{\{i+1jk\}})-n_z^{\{ijk\}})
                    \right]
    \Bigg]
=\\=
    \frac{1}{8}
        \bigg[
            C_r^{i} C_t^{j} C_r^{k}
                \left[
                    (n_x^{\{ijk+1\}}-n_x^{\{ijk\}})-(n_z^{\{i+1jk\}})-n_z^{\{ijk\}})
                \right]
            +\\+
            C_r^{i-1} C_t^{j} C_r^{k}
                \left[
                    (n_x^{\{i-1jk+1\}}-n_x^{\{i-1jk\}})-(n_z^{\{ijk\}})-n_z^{\{i-1jk\}})
                \right]
            +\\+
            C_r^{i} C_t^{j} C_r^{k-1}
                \left[
                    (n_x^{\{ijk\}}-n_x^{\{ijk-1\}})-(n_z^{\{i+1jk-1\}})-n_z^{\{ijk-1\}})
                \right]
            +\\+
            C_r^{i-1} C_t^{j} C_r^{k-1}
                \left[
                    (n_x^{\{i-1jk\}}-n_x^{\{i-1jk-1\}})-(n_z^{\{ijk-1\}})-n_z^{\{i-1jk-1\}})
                \right]
            +\\+
            C_r^{i} C_t^{j+1} C_r^{k}
                \left[
                    (n_x^{\{ij+1k+1\}}-n_x^{\{ij+1k\}})-(n_z^{\{i+1j+1k\}})-n_z^{\{ij+1k\}})
                \right]
            +\\+
            C_r^{i-1} C_t^{j+1} C_r^{k}
                \left[
                    (n_x^{\{i-1j+1k+1\}}-n_x^{\{i-1j+1k\}})-(n_z^{\{ij+1k\}})-n_z^{\{i-1j+1k\}})
                \right]
            +\\+
            C_r^{i} C_t^{j+1} C_r^{k-1}
                \left[
                    (n_x^{\{ij+1k-1\}}-n_x^{\{ij+1k-1\}})-(n_z^{\{i+1j+1k-1\}})-n_z^{\{ij+1k-1\}})
                \right]
            +\\+
            C_r^{i-1} C_t^{j+1} C_r^{k-1}
                \left[
                    (n_x^{\{i-1j+1k\}}-n_x^{\{i-1j+1k-1\}})-(n_z^{\{ij+1k-1\}}-n_z^{\{i-1j+1k-1\}})
                \right]
            \bigg]
    ,
\end{multline*}
\begin{multline*}
    \partial_{n^{\{ijk\}}_z}
    \Bigg[
        \frac{1}{8}
            \sum_{i=0}^{N_x-1} \sum_{j=0}^{N_y} \sum_{k=0}^{N_z-1}
                C_r^i C_t^j C_r^k
                \Big[
                    n_y^{\{ijk\}}+n_y^{\{i+1jk\}}+n_y^{\{ijk+1\}}+n_y^{\{i+1jk+1\}}
+\\+
                    n_y^{\{ij-1k\}}+n_y^{\{i+1j-1k\}}+n_y^{\{ij-1k+1\}}+n_y^{\{i+1j-1k+1\}}
                \Big]
                \left[
                    (n_x^{\{ijk+1\}}-n_x^{\{ijk\}})-(n_z^{\{i+1jk\}})-n_z^{\{ijk\}})
                \right]
    \Bigg]
=\\=
-
    \frac{1}{8}
        \bigg[
            C_r^{i-1} C_t^j C_r^k
                \left[
                    n_y^{\{i-1jk\}}+n_y^{\{ijk\}}+n_y^{\{i-1jk+1\}}+n_y^{\{ijk+1\}}+n_y^{\{i-1j-1k\}}+n_y^{\{ij-1k\}}+n_y^{\{i-1j-1k+1\}}+n_y^{\{ij-1k+1\}}
                \right]
-\\-
            C_r^i C_t^j C_r^k
                \left[
                    n_y^{\{ijk\}}+
                    n_y^{\{i+1jk\}}+
                    n_y^{\{ijk+1\}}+
                    n_y^{\{i+1jk+1\}}+
                    n_y^{\{ij-1k\}}+
                    n_y^{\{i+1j-1k\}}+
                    n_y^{\{ij-1k+1\}}+
                    n_y^{\{i+1j-1k+1\}}
                \right]
        \bigg]
    ,
\end{multline*}

\begin{multline*}
    \partial_{n^{\{ijk\}}_x}
    \Bigg[
        \frac{1}{8}
            \sum_{i=0}^{N_x-1} \sum_{j=0}^{N_y-1} \sum_{k=0}^{N_z}
                C_r^i C_r^j C_t^k
                    \Big[
                        n_z^{\{ijk\}}+n_z^{\{ij+1k\}}+n_z^{\{i+1jk\}}+n_z^{\{i+1j+1k\}}
+\\+
                        n_z^{\{ijk-1\}}+n_z^{\{ij+1k-1\}}+n_z^{\{i+1jk-1\}}+n_z^{\{i+1j+1k-1\}}
                    \Big]
                    \left[
                        (n_y^{\{i+1jk\}}-n_y^{\{ijk\}}) - (n_x^{\{ij+1k\}}-n_x^{\{ijk\}})
                    \right]
    \Bigg]
=\\=
    -
    \frac{1}{8}
    \bigg[
        C_r^i C_r^{j-1} C_t^k
            \Big[
                n_z^{\{ij-1k\}}+
                n_z^{\{ijk\}}+
                n_z^{\{i+1j-1k\}}+
                n_z^{\{i+1jk\}}+
                n_z^{\{ij-1k-1\}}+
                n_z^{\{ijk-1\}}+
                n_z^{\{i+1j-1k-1\}}+
                n_z^{\{i+1jk-1\}}
            \Big]
-\\-
        C_r^i C_r^j C_t^k
            \Big[
                n_z^{\{ijk\}}+n_z^{\{ij+1k\}}+n_z^{\{i+1jk\}}+n_z^{\{i+1j+1k\}}+n_z^{\{ijk-1\}}+n_z^{\{ij+1k-1\}}+n_z^{\{i+1jk-1\}}+n_z^{\{i+1j+1k-1\}}
            \Big]
    \bigg]
    ,
\end{multline*}

\begin{multline*}
    \partial_{n^{\{ijk\}}_y}
    \Bigg[
        \frac{1}{8}
            \sum_{i=0}^{N_x-1} \sum_{j=0}^{N_y-1} \sum_{k=0}^{N_z}
            C_r^i C_r^j C_t^k
                \Big[
                    n_z^{\{ijk\}}+
                    n_z^{\{ij+1k\}}+
                    n_z^{\{i+1jk\}}+
                    n_z^{\{i+1j+1k\}}
+\\+
                    n_z^{\{ijk-1\}}+
                    n_z^{\{ij+1k-1\}}+
                    n_z^{\{i+1jk-1\}}+
                    n_z^{\{i+1j+1k-1\}}
                \Big]
        \left[
            (n_y^{\{i+1jk\}}-n_y^{\{ijk\}}) - (n_x^{\{ij+1k\}}-n_x^{\{ijk\}})
        \right]
    \Bigg]
=\\=
    \frac{1}{8}
        \bigg[
            C_r^{i-1} C_r^j C_t^k
                \Big[
                    n_z^{\{i-1jk\}}+
                    n_z^{\{i-1j+1k\}}+
                    n_z^{\{ijk\}}+
                    n_z^{\{ij+1k\}}+
                    n_z^{\{i-1jk-1\}}+
                    n_z^{\{i-1j+1k-1\}}+
                    n_z^{\{ijk-1\}}+
                    n_z^{\{ij+1k-1\}}
                \Big]
-\\-
            C_r^i C_r^j C_t^k
                \left[
                    n_z^{\{ijk\}}+
                    n_z^{\{ij+1k\}}+
                    n_z^{\{i+1jk\}}+
                    n_z^{\{i+1j+1k\}}+
                    n_z^{\{ijk-1\}}+
                    n_z^{\{ij+1k-1\}}+
                    n_z^{\{i+1jk-1\}}+
                    n_z^{\{i+1j+1k-1\}}
                \right]
        \bigg],
\end{multline*}

\begin{multline*}
    \partial_{n^{\{ijk\}}_z}
    \Bigg[
        \frac{1}{8}
            \sum_{i=0}^{N_x-1} \sum_{j=0}^{N_y-1} \sum_{k=0}^{N_z}
                C_r^i C_r^j C_t^k
                    \Big[
                        n_z^{\{ijk\}}+n_z^{\{ij+1k\}}+n_z^{\{i+1jk\}}+n_z^{\{i+1j+1k\}}
+\\+
                        n_z^{\{ijk-1\}}+n_z^{\{ij+1k-1\}}+n_z^{\{i+1jk-1\}}+n_z^{\{i+1j+1k-1\}}
                    \Big]
                    \Big[
                        (n_y^{\{i+1jk\}}-n_y^{\{ijk\}})- (n_x^{\{ij+1k\}}-n_x^{\{ijk\}})
                    \Big]
    \Bigg]
=\\=
    \frac{1}{8}
    \bigg[
        C_r^{i} C_r^{j} C_t^{k}
            \left[
                (n_y^{\{i+1jk\}}-n_y^{\{ijk\}}) - (n_x^{\{ij+1k\}}-n_x^{\{ijk\}})
            \right]
+\\+
        C_r^{i} C_r^{j-1} C_t^{k}
            \left[
                (n_y^{\{i+1j-1k\}}-n_y^{\{ij-1k\}}) - (n_x^{\{ijk\}}-n_x^{\{ij-1k\}})
            \right]
+\\+
        C_r^{i-1} C_r^{j} C_t^{k}
            \left[
                (n_y^{\{ijk\}}-n_y^{\{i-1jk\}}) - (n_x^{\{i-1j+1k\}}-n_x^{\{i-1jk\}})
            \right]
+\\+
        C_r^{i-1} C_r^{j-1} C_t^{k}
            \left[
                (n_y^{\{ij-1k\}}-n_y^{\{i-1j-1k\}}) - (n_x^{\{i-1jk\}}-n_x^{\{i-1j-1k\}})
            \right]
+\\+
        C_r^{i} C_r^{j} C_t^{k+1}
            \left[
                (n_y^{\{i+1jk+1\}}-n_y^{\{ijk+1\}}) - (n_x^{\{ij+1k+1\}}-n_x^{\{ijk+1\}})
            \right]
+\\+
        C_r^{i} C_r^{j-1} C_t^{k+1}
            \left[
                (n_y^{\{i+1j-1k+1\}}-n_y^{\{ij-1k+1\}}) - (n_x^{\{ijk+1\}}-n_x^{\{ij-1k+1\}})
            \right]
+\\+
        C_r^{i-1} C_r^{j} C_t^{k+1}
            \left[
                (n_y^{\{ijk+1\}}-n_y^{\{i-1jk+1\}}) - (n_x^{\{i-1j+1k+1\}}-n_x^{\{i-1jk+1\}})
            \right]
+\\+
        C_r^{i-1} C_r^{j-1} C_t^{k+1}
            \left[
                (n_y^{\{ij-1k+1\}}-n_y^{\{i-1j-1k+1\}}) - (n_x^{\{i-1jk+1\}}-n_x^{\{i-1j-1k+1\}})
            \right]
    \bigg]
    .
\end{multline*}

(VI.4) Consider the gradient of $(\bn\cdot \bH)^2$ at a lattice site $\{ijk\}$: 
\begin{multline*}
    \partial_{\bn^{\{ijk\}}}
        \Bigg[
            \sum_{k=0}^{N_z}
                C_r^i C_t^j C_t^k
                    \left(
                        n_x^{\{ijk\}} H_x^{\{ijk\}}
                    \right)^2
            +
            \sum_{i=0}^{N_x} \sum_{j=0}^{N_y-1} \sum_{k=0}^{N_z}
                C_t^i C_r^j C_t^k
                    \left(
                        n_y^{\{ijk\}} H_y^{\{ijk\}}
                    \right)^2
            +
\\
            \sum_{i=0}^{N_x} \sum_{j=0}^{N_y} \sum_{k=0}^{N_z-1}
                C_t^i C_t^j C_r^k
                    \left(
                        n_z^{\{ijk\}} H_z^{\{ijk\}}
                    \right)^2
        \Bigg]
    =
    2\begin{bmatrix}
        C_r^i C_t^j C_t^k 
            \left(
                n_x^{\{ijk\}} H_x^{\{ijk\}}
            \right)
\\
        C_t^i C_r^j C_t^k
            \left(
                n_y^{\{ijk\}} H_y^{\{ijk\}}
            \right)
\\
        C_t^i C_t^j C_r^k
            \left(
                n_z^{\{ijk\}} H_z^{\{ijk\}}
            \right)
    \end{bmatrix}
    .
\end{multline*}



After substituting the expressions (V.1-V.4) into the equation  \eqref{eq:energy_lattice}, 
one can find the expression for the liquid crystal energy in the lattice approximation with a shift of the director components. 
Using the same equation and expressions (VI.1-VI.4) one can obtain an equation for the energy variation.

\section*{Acknowledgements}
This work was supported by the Foundation for the Advancement of Theoretical Physics and Mathematics ‘BASIS’ under Grant No. 19-1-1-12-5.

\bibliographystyle{plain}
\bibliography{tambovtsev}

\begin{thebibliography}{10}

\bibitem{bogi:lc:2001}
A~Bogi and Sandro Faetti.
\newblock Elastic, dielectric and optical constants of
  4'-pentyl-4-cyanobiphenyl.
\newblock {\em Liquid crystals}, 28(5):729--739, 2001.

\bibitem{bradshaw:mclc:1981}
MJ~Bradshaw and EP~Raynes.
\newblock Pre-transitional effects in the electric permittivity of cyano
  nematics.
\newblock {\em Molecular Crystals and Liquid Crystals}, 72(2-3):73--78, 1981.

\bibitem{cheng:jap:1981}
Julian Cheng, RN~Thurston, and DW~Berreman.
\newblock Boundary-layer model of field effects in a bistable liquid-crystal
  geometry.
\newblock {\em Journal of Applied Physics}, 52(4):2756--2765, 1981.

\bibitem{degennes:book:1993}
de~Gennes and P.~G.;~Prost J.
\newblock {\em The Physics of Liquid Crystals}.
\newblock Clarendon Press, Oxford, UK, 1993.

\bibitem{deuling:mclc:1972deformation}
Heinz~J Deuling.
\newblock Deformation of nematic liquid crystals in an electric field.
\newblock {\em Molecular Crystals and Liquid Crystals}, 19(2):123--131, 1972.

\bibitem{hakemi:jcp:1983}
H~Hakemi, EF~Jagodzinski, and DB~DuPre.
\newblock The determination of the elastic constants of a series of
  n-alkylcyanobiphenyls by anisotropy of turbidity.
\newblock {\em The Journal of Chemical Physics}, 78(3):1513--1518, 1983.

\bibitem{hara:mclc:1985}
Masahiko Hara, Jun-Ichi Hirakata, Takehiro Toyooka, Hideo Takezoe, and Atsuo
  Fukuda.
\newblock Determination of the frank elastic constant ratios in nematic liquid
  crystals (ncb) by observing angular dependence of rayleigh light scattering
  intensity.
\newblock {\em Molecular Crystals and Liquid Crystals}, 122(1):161--168, 1985.

\bibitem{knyazev:lc:2013}
Andrey~A Knyazev, Elena~Yu Molostova, Aleksandr~S Krupin, B~Heinrich, B~Donnio,
  Wolfgang Haase, and Yury~G Galyametdinov.
\newblock Mesomorphic behaviour and luminescent properties of
  mesogenic-diketonate lanthanide adducts with 5, 5'-di (heptadecyl)-2,
  2'-bipyridine.
\newblock {\em Liquid Crystals}, 40(7):857--863, 2013.

\bibitem{oswald:pr:2000}
P~Oswald, J~Baudry, and S~Pirkl.
\newblock Static and dynamic properties of cholesteric fingers in electric
  field.
\newblock {\em Physics Reports}, 337(1-2):67--96, 2000.

\bibitem{rybakov:prl:2015}
F.~N. Rybakov, A.~B. Borisov, S.~Bl\"ugel, and N.~S. Kiselev.
\newblock New type of stable particlelike states in chiral magnets.
\newblock {\em Physical Review Letters}, 115:117201, 2015.

\bibitem{srivastava:jpcm:2004}
Amit Srivastava and Shri Singh.
\newblock Elastic constants of nematic liquid crystals of uniaxial symmetry.
\newblock {\em Journal of Physics: Condensed Matter}, 16(41):7169, 2004.

\bibitem{tai:pre:2020:smalyukh}
Jung-Shen~B Tai, Ivan~I Smalyukh, et~al.
\newblock Surface anchoring as a control parameter for stabilizing torons,
  skyrmions, twisted walls, fingers, and their hybrids in chiral nematics.
\newblock {\em Physical Review E}, 101(4):042702, 2020.

\end{thebibliography}

\end{document}